\documentclass[12pt]{article}
\usepackage[dvips]{graphicx}

\title{
\hspace{10 cm}{\small ISU-IAP.TH 99-01,Irkutsk.}\\
\vspace{1 cm}
The polarization of a final electron
in the field of intensive electromagnetic wave\\
}
\author{E.M.Bol'shedvorsky and S.I.Polityko \\
Irkutsk State University\\
E-mail:polityko@psi.isu.ru
}
\date{1999}

\begin{document}

\maketitle
\begin{center}
{\bf Abstract}
\end{center}

     The polarization properties of the Compton effect in the field
of a circularly polarized electromag\-netic wave are discussed.
In the case of nonlinear Compton effect the behavior of spin is analyzed.
The transformation of the longitudinal polarization in the transverse
polarization is showed. We obtained the total formulas describing the
polarization of the final electron in the nonlinear case.

\vspace{0.5cm}
\centerline { \bf 1.Introduction }
\vspace{0.5cm}
        The Compton effect is the main process obtaining from intensive
high energy $\gamma $ bunches on the base of the present linear electron
collider, so named electron-photon and photon-photon collideres (PLC)
\cite{1},\cite{Tel1}. The polarization properties of this process are
very important. For example, if the helities of initial particles are
opposite the high energy part of spectrum increased twice comparing
with the nonpolarizing case.

The optimal choice of the conversion coefficient is important  in the conversion
scheme (the relation between
hard photon number to number of electron in bunches).
The conversion coefficient is proportional to the laser flash energy
and  its increasing in  the conversion region causes the important processes
of multiple
scattering electrons on a laser photon .
In this case the effects of nonlinear quantum electrodinamics are
essential \cite{NQED}.
In this study we consider the polarization of final electron in polarized
electron scattering on a circularly polarized laser photons via absorption
of $s$ photons from a laser wave and perform summation over the
final-photon  polarizations. Some particular cases in the problem
of final electron polarization in nonlinear QED were treated in \cite{DP}
but the leading formulas can be used only in the c.m.s. of colliding particles.
However, $e \rightarrow \gamma $ - conversion must be considered in a
different frame where an electron with energy $E\sim 100$ Ē'
collides with a laser photon with energy $\omega \sim 1$ í'.
This frame will be referred  as a collider frame. The linear case of this
problem was considered earlier in \cite{KPSNIM}.

 We shall consider the process of photon emission by an electron
 via simultaneous absorption of photons from a laser wave
 (nonlinear Compton effect).

   $$
 s\gamma _0 + e^- \rightarrow \gamma + e^- \eqno (1)
   $$

       The systematic investigation of such a process was performed in
\cite{Ritus},\cite{LD}.
The possibility of observing nonlinear QED effects for an unpolarized
electron beam in reaction (1) was discussed earlier in \cite{NQED}.
It was shown in \cite{Tel1},\cite{GKPS} that the application of
longitudinally polarized electron beams and circularly polarized laser
light leads to substantial improvements in the monochromaticity of
$\gamma \gamma $ collisions. Anyway probability, there are the initial-beam
polarizations that will be used in the implementation of concepts based on
colliding $\gamma e$ and $\gamma \gamma $ beams.

However, when we investigate the Compton scattering
of a high energy electron by a laser beam ($e\to \gamma$ conversion)
at a comparatively large conversion ratio, we must take into
consideration multiple electron scattering by laser photons.
Because the required formulas were not available the similar calculations
have already been performed in papers \cite{BBC},
but they either considered the conversion of unpolarized particles or
disregarded the change in electron polarization in the scattering process
.
However, it is necessary to know the polarization of the final electron,
since it is the initial electron for the next scattering event.

In this paper we calculate the polarization of the final electron
and discuss its characteristics features.
It is found that polarization effects are essential and that they must be
taken into account in calculations dealing with multiple scattering.
\vspace{0.5cm}
\begin{center}
      {\bf 2. The bases for polarization describing\\
 of initial and final particles}
\end{center}
\vspace{0.5cm}
Nonlinear QED processes are characterized by the parameter
$$
\xi ^2 = {e^2F^2\hbar^2\over m^2\omega_0^2c^2},
$$
\par
\noindent
where $F$-is the strength of the electromagnetic-wave field,
$m$-is the mass of electron,
$\omega_0$ is the energy of laser photon. At $\xi^2 \ll 1$ an electron
interacts only with one photon (usual Compton effect),
at $\xi^2 \gg 1$ an electron interacts with a collective field
(synchrotron radiation). The transverse motion of electron
in the electromagnetic-wave field
leads to increase the effective electron mass
$m_*^2=m^2(1+\xi^2)$ and changes the spectrum of scattering photons:
the high energy part of the spectrum with $s=1$ squeezes and
the high energy tail appears, corresponding to the absorption from
the wave two or more laser photons.
Considering that the electron in the electromagnetic-wave field
becomes "heavier" we can write the energy-momentum conservation law
in the form
$$
q_1+sk_1=
q_2+k_2,
$$
\par
\noindent where $q_1,q_2$ "quasimomentums" of initial and final electrons
$$
q_1=p_1+ \xi^{2}{m^2 \over 2(k_1p_1)}k_1,\quad \quad
q_2=p_2+ \xi^{2}{m^2 \over 2(k_1p_2)}k_1.
$$
Let us introduce the invariants for describing of scattering process:

$$
x={2p_1k_1\over m^2}, \;\;\; y=\frac{k_1k_2}{p_1k_1}.
$$
Let {\boldmath $\zeta$} and {\boldmath $\zeta ^{\prime}$} ---
be the polarization vectors  of the initial and final electrons.
They determine the electron-spin 4-vectors as
$$
\sigma=\left({\mbox {\boldmath $\zeta$}{\bf p_1}\over m}\,,\,
\mbox{\boldmath $\zeta$}+ {\bf p_1}{\mbox{\boldmath $\zeta$} {\bf
p_1} \over m(E_1+m)}\right) \,,\;\;\;
\sigma^{\prime}=\left({\mbox {\boldmath$\zeta^{\prime}$} {\bf
p_2}\over m}\,,\, \mbox {\boldmath$\zeta^{\prime}$} +{\bf
p_2}{\mbox {\boldmath$\zeta^{\prime}$} {\bf
p_2}\over m(E_2+m)}\right)
$$
and the average helicities of the initial and final electrons as
$$
\lambda_e={\mbox{\boldmath $\zeta$}{\bf p_1}\over 2|{\bf
p_1}|}\,,\;\;\;\;
\lambda^{\prime}_e={\mbox {\boldmath$ \zeta^{\prime}$} {\bf
p_2}\over 2|{\bf p_2}|}\,.
$$
In the collider system we choose the $z$ axis along the initial
electron momentum. We denote by $\theta$ and $\varphi$ the polar and
azimuthal angles of final photon emission. All azimuthal angles are defined
with respect to the fixed $x$ axis.
In this frame, the expression to which the invariants
are reduced for $\theta \ll 1$ are given by
$$
x={4E_1\omega_1\over m^2}\,,\;\;\;\; y={\omega_2\over
E} = 1- {E_2\over E_1};
$$
$$
\zeta_1 =\mbox{\boldmath$\zeta$}\;{[{\bf k_1},{\bf k_2}]\over|[{\bf
k_1},{\bf k_2}] |} =\zeta_{\perp} \sin{(\varphi-\beta)},\;\;
$$
$$
\zeta_2 = \mbox{\boldmath$\zeta$} \;{{\bf k_2}_{\perp}\over|{\bf
k_2}_{\perp}|}  =\zeta_{\perp}
\cos{(\varphi-\beta)} ,\;\; \zeta_3 =2\lambda_e,
$$
where $\omega_2$ and $E_2$ are the energies of the final photon and
electron,
${\bf k_2}_{\perp}$ and $\mbox{\boldmath$\zeta$} _{\perp}$
are the transverse components of the vectors ${\bf k_2}$ and $\mbox{\boldmath
$\zeta$}$ with respect to the vector ${\bf q}$,  $\varphi$ and $\beta$
are the azimuthal angles of ${\bf k_2}$ and $\mbox{\boldmath$\zeta$}; \;\;
\zeta_{\perp} = |\mbox{\boldmath$\zeta$} _{\perp}|$.

Thus, $\zeta_1$ is the initial-electron transverse polarization
that is orthogonal to the scattering plane, $\zeta_2$ is the transverse
polarization lying in this plane.

We decompose the spin 4-vectors on the orthogonal system of bases vectors
The bases vectors, describing the polarization state of initial particles
are given
$$
e_0=\frac{p_1}{m},\quad
e_1=n_3,
$$
$$
e_2=-\frac{z}{2\xi\sqrt{sxy}}x(1-y)(n_0-n_1)-n_2,\quad
e_3=\frac{1}{m}(p_1-\frac{2}{x}k).
$$
For the final particles the basic vectors have the form
$$
e'_0=\frac{p_2}{m},\quad
e'_1=n_3,
$$
$$
e'_2=-\frac{z}{2\xi\sqrt{sxy}}x(n_0-n_1)-n_2,\quad
e'_3=\frac{1}{m}(p_2-\frac{2}{x(1-y)}k).
$$
In this expressions
$$
n_0=\frac{K}{m\sqrt{sxy}},\quad
n_1=\frac{q}{m\sqrt{sxy}},\quad
n_2=\frac{\xi y}{mzx(1-y)}P\bot,
$$
$$
n_3^{\alpha}=\frac{\xi }{sm^3zx^2(1-y)}
e^{\alpha \mu \rho \nu}K_{\mu}q_{\rho}P_{\nu},\quad
$$
\par
\noindent
where
$$
K=sk_1+k_2,\quad
q=k_2-sk_1=q_1-q_2,\quad
P\bot=P-\frac{PK}{K^2}K,\quad
P=q_1+q_2,
$$
$$
z=\frac{2s\xi}{\sqrt{1+\xi^2}}\sqrt{r(1-r)},\quad
r={y(1+\xi^2)\over (1-y)sx}.
$$
These bases differ from the bases chosen in  \cite{KPSNIM},
here we have made a turn  around the axes $z^\prime$
at the angle $\theta^{*}$ which is considered to be a scattering angle
of a photon in the rest frame of initial electron
$$\cos\theta^{*}=1-\frac{2y}{sx(1-y)-\xi^2y}.$$
We can represent the 4-vectors of spin $\sigma$ and $\sigma^{\prime}$ as
$$
\sigma=\sum_j \zeta_j e_j\; ,\;\;\; \zeta_j=-\sigma e_j, \;,\;\;\;
\qquad
\sigma^{\prime}=\sum_j z^{\prime}_j e^{\prime}_j \;,\;\;\;
z^{\prime}_j=-\sigma^{\prime}e^{\prime}_j\, .
$$
In the collider frame $\zeta^\prime$ mean
$$
\zeta'_1=\zeta'_{\perp}\sin(\varphi-\beta'),\quad
\zeta'_2=\zeta'_{\perp}\cos(\varphi-\beta'),\quad
\zeta'_3=2\lambda^\prime_e,
$$
where $\zeta'_{\perp}$ is the transverse component of the vector
${\bf \zeta'}$ with respect to the vector ${\bf p_2}$,
and $\beta'$ is the azimuthal angle for the direction ${\bf\zeta'}$.
Thus $\zeta^\prime_3$ is the double helicity of final electron,
$\zeta'_2$ is the transverse polarization  lying in scattering plane,
and $\zeta'_1$ is orthogonal to this plane.The invariants
$\zeta_j$ and $z^{\prime}_j$ completely characterize the polarization
properties of electrons.

\vspace{0.5cm}
\centerline{\bf 3. Differential probability of emission}
\vspace{0.5cm}

        We will calculate the probability of the process (1) using the method
of quantum transition.As for the basis we use the exact nonstationary
solutions of the Dirac equation (Volkov's solutions ).
We take into account the electron interaction with the external field .
Electron interaction with the field of emitted photons is considered
with the help of  perturbation theory.

        The amplitude of the electron transition from the state $\psi _{p_1}$
in the wave field to the state $\psi _{p_2}$ with the emission of a photon
of momentum $k_2$ and polarization $\varepsilon _2 $ is given by

$$
M=e\int \bar \Psi _{p_2}(x)\hat \varepsilon _2^* \Psi_{p_1}(x)
{e^{-ik_2x}\over \sqrt {2E_2}} d^4x.\eqno(2)
$$
\par
\noindent
where $\varepsilon _2$ is the final-photon polarization,
$\Psi _{p_1}(x) \quad (\bar \Psi _{p_2}(x))$ is the exact solution
of the Dirac equation for an electron in the field of a circularly
polarized wave :
$$
\Psi _{p_1}(x)=(1+{ek_1\hat A \over 2k_1p_1})u_{p_1}\exp
(ie{a_{1}p_1 \over k_1p_1} \sin (\varphi )-ie{a_{2}p_1 \over k_1p_1} \cos
(\varphi )
+iqx),\quad \varphi =k_1x,
$$
\par
\noindent
The vector potential of a circularly polarized electromagnetic wave
has the form
$$
A=a_1\cos \varphi +P_ca_2 \sin \varphi,
$$
\par
\noindent
where $k_1a_1=k_1a_2=a_1a_2=0$,$\quad a_1^2=a_2^2=a^2$.
The average helicity of final (laser) photon $P_c$ coincide with
the Stokes parameter $\xi_2$.
The density matrices of the polarized electrons are given by the standard
form
$$
\rho=\frac{1}{2}(\hat p_1+1)(1-\gamma^5\hat \sigma),\quad
\rho'=\frac{1}{2}(\hat p_2+1)(1-\gamma^5\hat \sigma^\prime),
$$
\par
\noindent
where $\sigma,\sigma^\prime$ are 4-vectors of electron spin.

The density matrice of a final photon is
$$
\nu_{ik}=e'_ie'^{*}_k=-\frac{g_{ik}}{2}
$$

        Using the standard techniques developed in \cite{Ritus},\cite{LD}
we can easily calculate the probability of process (1) and perform summation
on final-photon polarizations.
$$
dW_s=\frac{\alpha m^2}{2E_1}\xi^2d\varphi dy\biggl(F_{00}^{00}+
P_c\sum_{i=1}^{3}(F^{00}_{i2}\zeta_i+F_{02}^{i0}\zeta'_i)
+\sum_{i=1}^{3}\sum_{j=1}^{3}F_{i0}^{j0}\zeta_i\zeta'_j\biggr),\eqno(3)
$$
\par
\noindent
where $\alpha=e^2/4\pi $. Nonzero coefficients $F_{ik}^{j0}$ are given
$$
F^{00}_{00}=-\frac{2}{\xi^2}J^2_s(z)+\biggl(1+\frac{y^2}{2(1-
y)}\biggr)(J^2_{s+1}(z)+J^2_{s-1}(z)-2J^2_s(z))\\
$$
$$
F^{00}_{32}=F^{30}_{02}=J_s(z)J'_s(z)\frac{s(2-
y)}{z}\biggl(\frac{z^2x}{\xi^2s}-2\frac{y}{1-y}\biggr)\\
$$
$$
F^{00}_{22}=-2J_{s}(z)J'_{s}(z)\frac{y}{\xi}\\
$$
$$
F^{20}_{02}=-2J_s(z)J'_s(z)\frac{y}{\xi(1-y)}\eqno(4)
$$
$$
F^{30}_{30}=J_s^2(z)\frac{1}{\xi^2}\biggl(2-(2+\xi^2)\frac{y^2-2y+2}{1-
y}\biggr)+\frac{1}{2}\frac{y^2-2y+2}{1-y}(J_{s+1}^2(z)+J_{s-1}^2(z))\\
$$
$$
F^{30}_{20}=\biggl(-\frac{2s}{z}\frac{y}{\xi}+\frac{z}{\xi^3}x(1-
y)\biggr)J_s^2(z)\\
$$
$$
F^{20}_{30}=\biggl(\frac{2s}{z}\frac{y}{\xi(1-y)}-
\frac{z}{\xi^3}x\biggr)J_s^2(z)\\
$$
$$
F^{20}_{20}=F^{10}_{10}=\biggl(J_{s+1}^2(z)+J_{s-1}^2(z)\biggr)-
2\biggl(\frac{1}{\xi^2}+1\biggr)J_s^2(z)
,
$$
\par
\noindent
where $J_s(z)$ is Bessel function.

\vspace{0.5cm}

\centerline{\bf 4. Numerical calculations}
\vspace{0.5cm}

        For numerical calculations the physical parameters characterizing
the region of laser conversion are chosen in accordance with \cite{Tel2}.
In the conversion scheme it has been proposed to use a solid-state
laser with laser photon energy $\omega _{0}=1.17$ í'. Let us choose
the parameter $x=4.8$. Which is according to the threshold of $e^+e^-$ pair
production at collision of   laser photon with  high energy scattering photon.
The length of the conversion region characterized by a high density of
laser photon $l_\gamma$ is close to the length of the electron bunch $l_e$.
It is convenient to express the parameter $\xi ^2$ in terms of the energy
$A$, the duration $\tau $ and the radius $a_{\gamma }$ of the laser beam
at the point of intersection

$$
\xi ^{2} = {A \over A_{*}},\quad {\rm where}\quad A_{*} =
{\tau c \over 4} \cdot \left ({ m_{e} \omega _{0} a_{\gamma} c \over
e \hbar } \right ) ^{2}.
$$
        The conversion coefficient $k$ connect with the laser flash energy
by relation
$$
k=N_\gamma /N_e \sim 1-\exp(-A/A_*)\sim A/A_* ,
$$
\par
\noindent
thus at $x=4.8$ and $k=1$ we get $\xi^2 =0.05E[TeV]/l_\gamma$\cite{Tel2}.
If $l_\gamma =200\mu m$  and $E=100$ Gev, then $\xi^2 =0.25$.
The expression (3) for the differential probability can be written
in the following way
$$
dW_{s}=\frac{\alpha m^2}{2E_1}\xi^2d\varphi dy
\biggl(G_0+\sum_{i=1}^{3}G_{i}\zeta'_{i}\biggr),
$$
\par
\noindent
where
$$
G_{0}=F_{00}^{00}+P_cF_{32}^{00}\zeta_{3}+P_cF_{22}^{00}\zeta_{2},
$$
$$
G_{1}=F_{10}^{10}\zeta_1,
$$
$$
G_{2}=P_cF_{02}^{20}+F_{30}^{20}\zeta_3+F_{20}^{20}\zeta_2,
$$
$$
G_{3}=P_cF_{02}^{30}+F_{30}^{30}\zeta_3+F_{20}^{30}\zeta_2.
$$
The components of the polarization vector of final electron
are given
$$
\zeta'^f_1=\frac{G_1}{G_0},\quad
\zeta'^f_2=\frac{G_2}{G_0},\quad
\zeta'^f_3=\frac{G_3}{G_0}. \eqno(5)
$$
       These formulas determine completely the polarization of final electron
in the field of intensive electromagnetic wave. The case $\xi \gg 1$ is
described in the Appendix.

       In figures 1-6 the numerical calculations
 present the components of the polarization vector of final electron
dependence $\zeta^{\prime (f)}_i$ as functions of its energy $y_1=E^\prime/E
= 1-y$ for various degrees of polarization of the initial electron and photon
at $x=4.8$.
The number of absorbed laser photons $s=5$  and the parameter $\xi^2 =0.25$.
Here $\zeta^{\prime (f)}_1$ is the final electron transverse polarization
in the direction orthogonal to the scattering plane,
$\zeta^{\prime (f)}_2$ is the transverse polarization in this plane
and $\zeta^{\prime (f)}_3$ is the double average helicity.
The analogous quantities for the initial electron are denoted by
 $\zeta_1,\;\zeta_2$ ¨ $\zeta_3=2 \lambda_e$; $P_c$ is the average
helicity of initial photon. For the figures 1-6 we indicate nonzero
components of the initial particles, for the figures 1-5
$\zeta_1^{\prime (f)}=0$.

  The appearance of the essential transverse polarization
on the scattered electron is the common property shown in the figures.
The transverse polarization takes place even in the case when the initial
electron is not polarized.  The electron polarization in the scattering plane
changes the sign by altering the laser photon helicity (see.fig.1b and 2b).
Hard scattering electrons for which $y_1 \approx 1$ conserve their
helicity. In the soft region (small $y_1$) scattered electrons get the longitudional
polarizations its sign depends on laser photon polarization sign.
In this region the multiple photon absorption is essential.

We note, that at  $\xi^2 \ll 1$ and $s=1$ we get the results of\cite{KPSNIM}.

We are very grateful to I.~Ginzburg, G.~Kotkin and V.~Serbo for
useful discussions.

\vspace{0.5cm}
\centerline{\bf Appendix}
\vspace{0.5cm}

In the Appendix we shall consider the formulas for the probability of photon
radiation by electron at large values $ \xi $. The large value $ \xi$
can be obtained from decreasing the frequency at  fixed field strength.

With increasing $ \xi$ the energy of wave the quant decreases but the number of
absorption photons from electromagnetic wave increases $s\sim \xi^3$ and
the mean energy and momentum of final electron increase effectively.
The values are
$$ x^{\ast}=\frac{2k_1p_1}{m^2}\xi,\quad y,\quad \rho_1=
\frac{\alpha_1}{8\beta},\quad \rho_2=\frac{\alpha_2}{8\beta},
$$
where
$$\beta =-\frac{\xi^2m^2}{8}\Big(\frac{1}{k_1p_1}-
\frac{1}{k_1p_2}\Big)=\frac{\xi^2y}{4x^{\ast}(1-y)},
$$
does not depend on $\xi$.
The variables $\rho_{1,2}$ are connected with azimuthal angle $\varphi_0$ as
$$
\rho_1=\frac{z}{8\beta}\cos \varphi_0,\quad
\rho_2=\frac{z}{8\beta}\sin \varphi_0,
$$
where
$$\cos\varphi_0=\frac{\alpha_1}{z},\quad\sin\varphi_0=\frac{\alpha_2}{z}.$$
Note that at $z\sim \xi^3$ the value $z/8\beta$ leads to unit effectively

We describe the difference of $z/8\beta$ from the unit as the new variable $\tau$:
$$
z=8\beta \sqrt{1+\frac{2\tau}{\xi}}=\frac{2\xi^3y}{x^{\ast}(1-y)}
\sqrt{1+\frac{2\tau}{\xi}}.
$$
Let's consider the probability of photon radiation by electron. Changing the
order of integration and summation in the expression of the probability,
we get
$$ W=\sum_{s=1}^{\infty}\int_0^{2\pi}d\varphi \int_0^1dyw=
\int_0^{2\pi}d\varphi \int_0^1dy\sum_{s>s_{min}
}^{\infty}w,$$
where
$$s_{min}=\frac{\xi^3y}{x^{\ast}(1-y)}\left(1+\frac{1}{\xi^2}\right).$$

At $\xi \gg 1$ the large number of terms and the sum can be changed by the
integral.
From the dependence $z$ and$s$ it follows that
$$s=\frac{\xi^3y}{x^{\ast}(1-y)}\left(1+\frac{2\tau}{\xi}\right)+s_{min},$$

Then
$$W=\int_{0}^{2\pi}d\varphi \int_0^1dy\int_{-\xi^2/2}^{\infty}d\tau \frac{2\xi^2y}{x^{\ast}(1-y)}w.$$

At  $\xi \gg 1$ the value $z\sim s\sim \xi^3$, but their relation leads to
the unit,
$$\sigma =\xi^2\left(1-\frac{z^2}{s^2}\right)=1+\tau^2,$$
and the function Bessel can be changed by the asymptotic Vatson
representation
$$J_s(z)\approx \frac{1}{\pi}\left(\frac{2}{s}\right)^{1/3}Ai(\theta),
\quad \theta =\left(\frac{s}{2}\right)^{2/3}\left(1-
\frac{z^2}{s^2}\right)=\left(\frac{y}{x^{\ast}(1-y)}\right)^{2/3},
$$
where $Ai(\theta)$ is the Eiri function.

Using this representation we finally get
$$
W(\infty,x^{\ast})=2\frac{\alpha m^2}{E_1\pi^2}\int_0^{2\pi}d\varphi
\int_0^1dy\int_{-\infty}^{\infty}d\tau \left(\frac
{\theta}{\sigma}\right)^{1/2}\left( F_{00}^{00}+P_c\sum_{i=1}^3
\left( F_{i2}^{00}\zeta_i+F_{02}^{i0}
\zeta_i'\right)+\sum_{i=1}^3\sum_{j=1}^3F_{i0}^{j0}\zeta_i\zeta_j'\right),
$$
where
$$F_{00}^{00}= -Ai^2(\theta)+\frac{\sigma}{\theta}(1+\frac{y^2}{2(1-y)})
\left(\theta Ai^2(\theta)+\acute{Ai}^2(\theta)\right) $$
$$F_{32}^{00}=F_{02}^{30}=-\left(\frac{\sigma}{\theta}\right)^{1/2}\frac{y(2-y)}{1-y}Ai(\theta)
\acute{Ai}(\theta)\tau$$
$$F_{22}^{00}=\left(\frac{\sigma}{\theta}\right)^{1/2}Ai(\theta)\acute{Ai}(\theta)y$$
$$F_{02}^{20}=\left(\frac{\sigma}{\theta}\right)^{1/2}Ai(\theta)\acute{Ai}(\theta)\frac{y}{1-y}$$
$$F_{30}^{30}=\frac{y-y^2-1}{1-y}Ai^2(\theta)+\frac{1}{2}\frac{y^2-2y+2}{1-y}\frac{\sigma}{
\theta}\left(\theta Ai^2(\theta)+\acute{Ai}^2(\theta)\right)$$
$$F_{20}^{30}=Ai^2(\theta)y\tau$$
$$F_{30}^{20}=Ai^2(\theta)\frac{y}{1-y}\tau$$
$$F_{20}^{20}=F_{10}^{10}= -Ai^2(\theta)+\frac{\sigma}{\theta}
\left(\theta Ai^2(\theta)+\acute{Ai}^2(\theta)\right) $$

The polarization of final electron is determined by formulae (5),
where
$$
G_0=\int_{-\infty}^{\infty}d\tau\left(F_{00}^{00}+P_cF_{32}^{00}\zeta_3+P_c
F_{22}^{00}\zeta_2\right),
$$
$$
G_1=\int_{-\infty}^{\infty}d\tau \left(F_{10}^{10}\zeta_1\right),
$$
$$
G_2=\int_{-\infty}^{\infty}d\tau \left(P_cF_{02}^{20}+F_{30}^{20}\zeta_3+
F_{20}^{20}\zeta_2\right),
$$
$$
G_3=\int_{-\infty}^{\infty}d\tau \left(P_cF_{02}^{03}+F_{30}^{30}\zeta_3+
F_{20}^{30}\zeta_2\right).
$$

\newpage
\vspace{1cm}
\begin{figure}
\hspace{2cm}
\includegraphics[bb=110 630 240 770,scale=0.3]{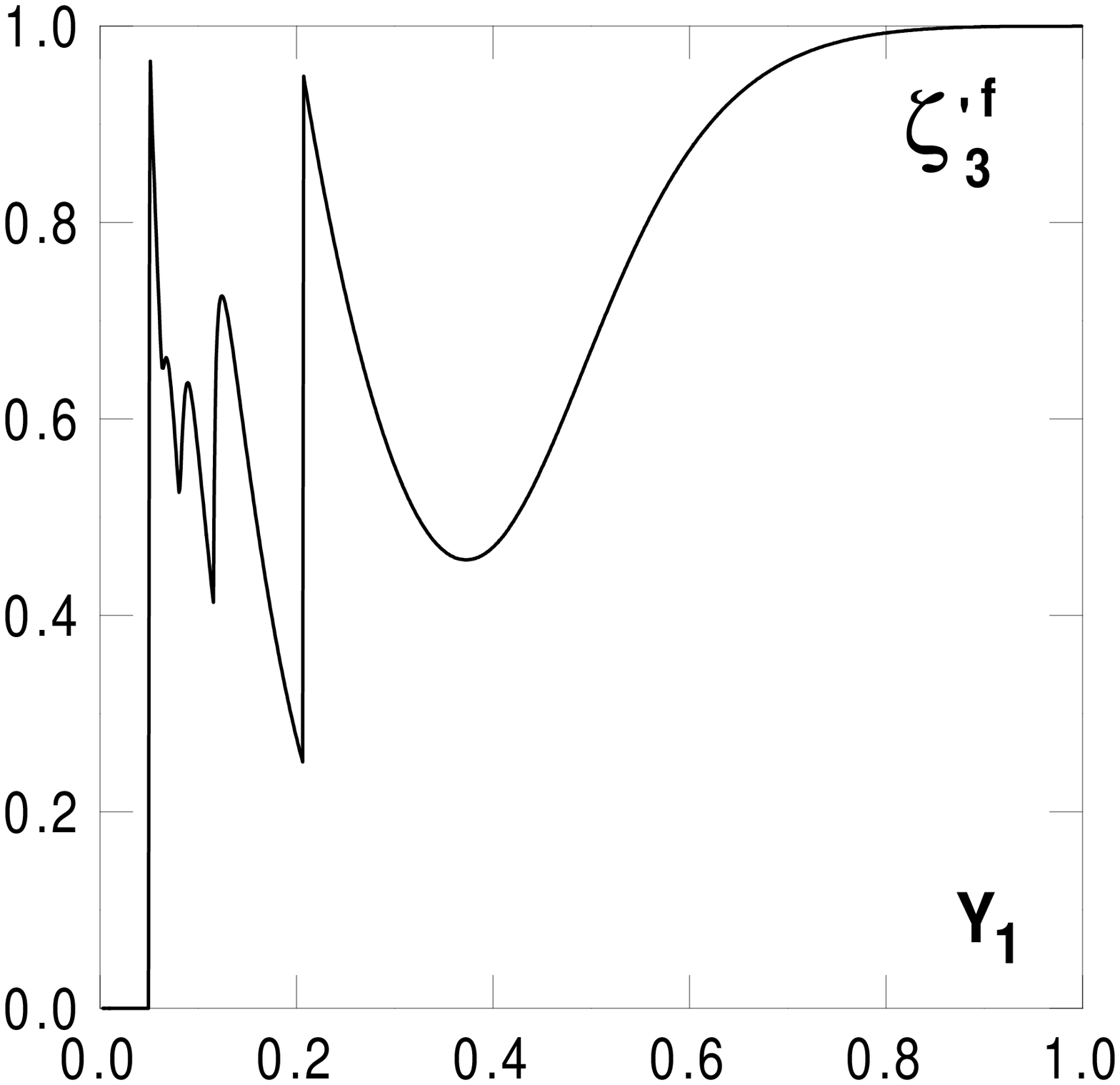}
\hspace{7.5cm}
\includegraphics[bb=340 630 480 770,scale=0.3]{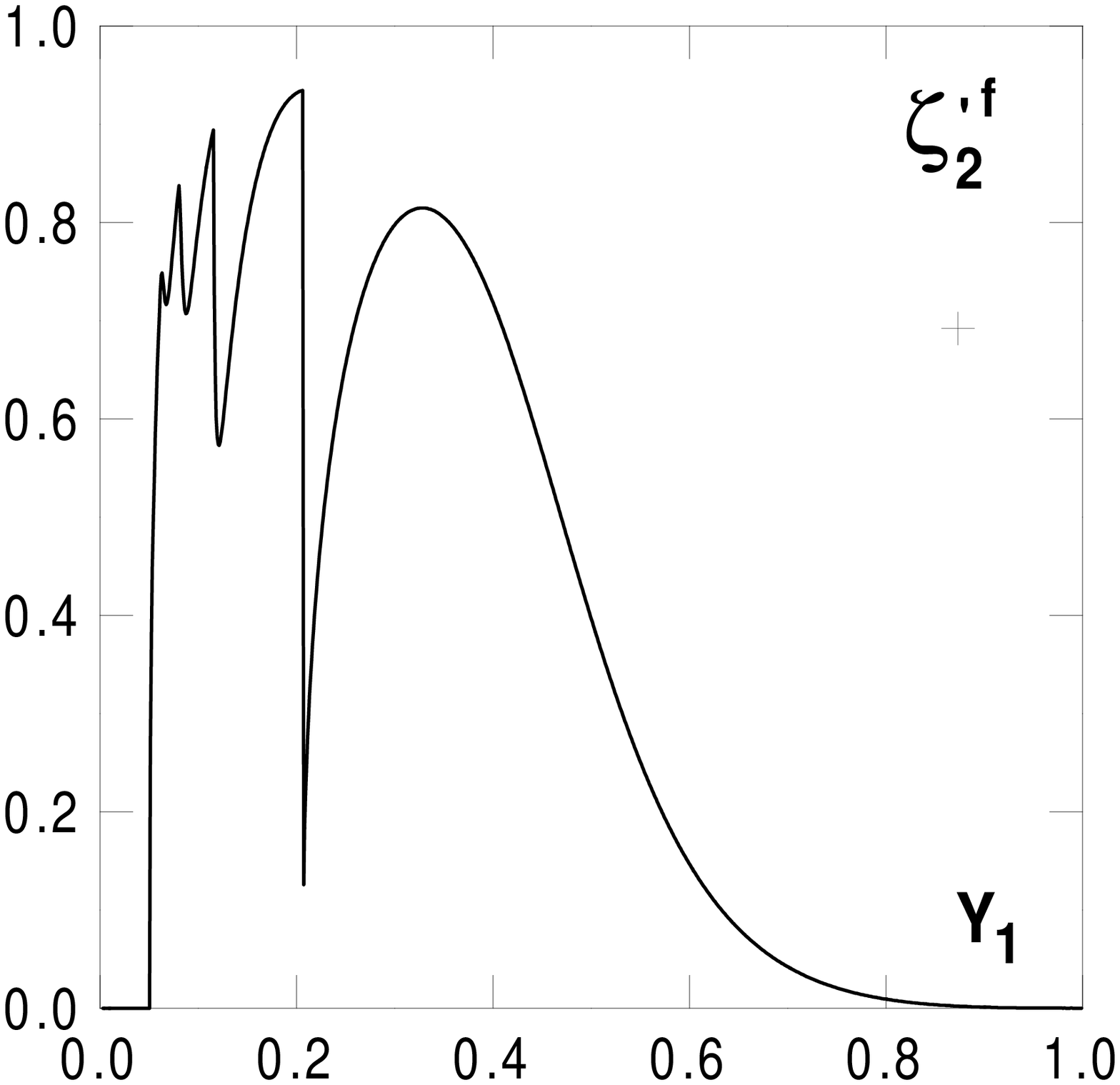}
\vspace{5cm}
\caption{Components of the polarization vector of the scattered electron
of its energy
$\zeta^{\prime (f)}_2$, $\zeta^{\prime (f)}_3$ at $2\lambda_e=1,\; P_c=-
1$;}
\end{figure}

\begin{figure}
\hspace{2cm}
\includegraphics[bb=110 630 240 770,scale=0.3]{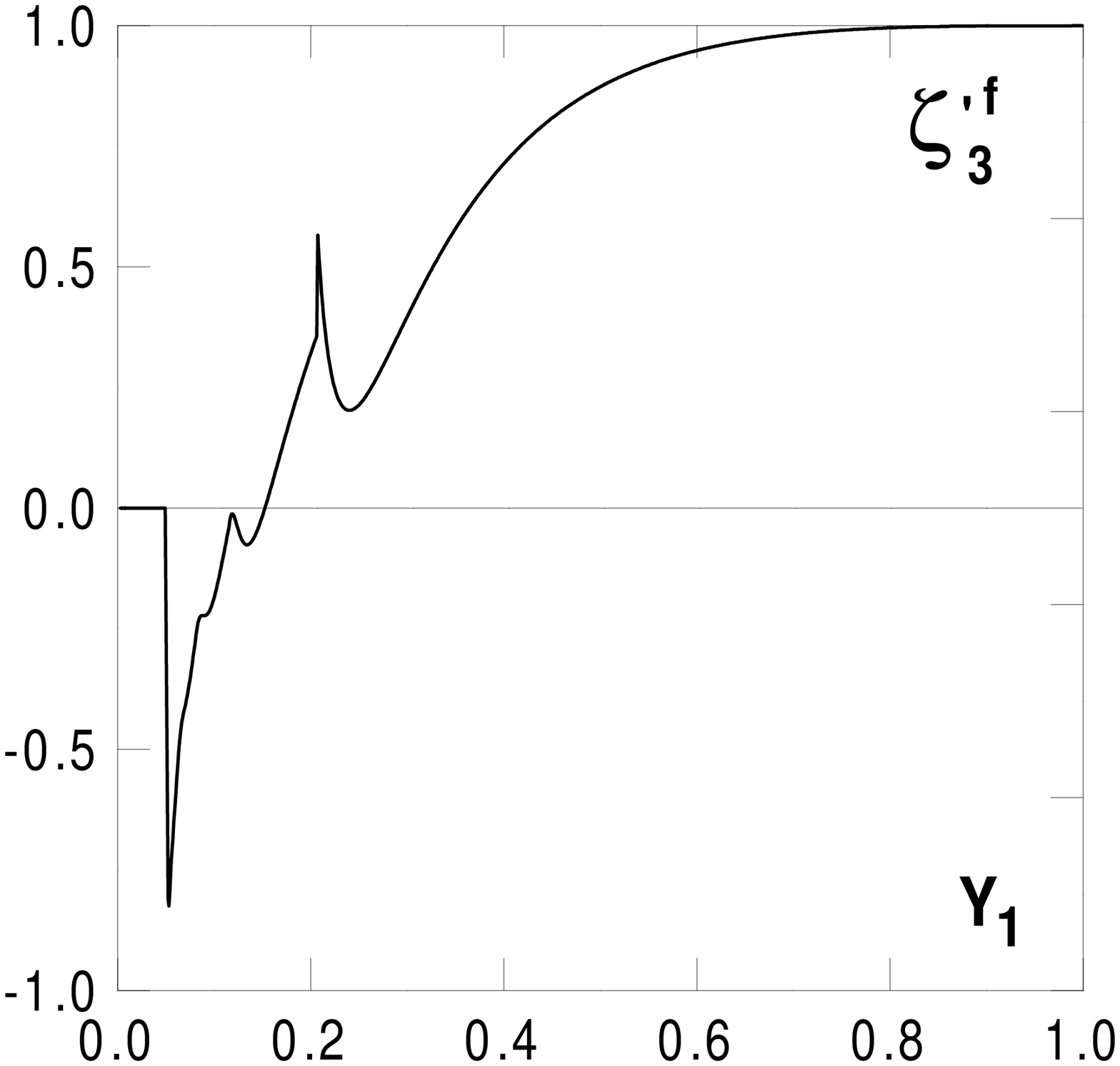}
\hspace{7.5cm}
\includegraphics[bb=340 630 480 770,scale=0.3]{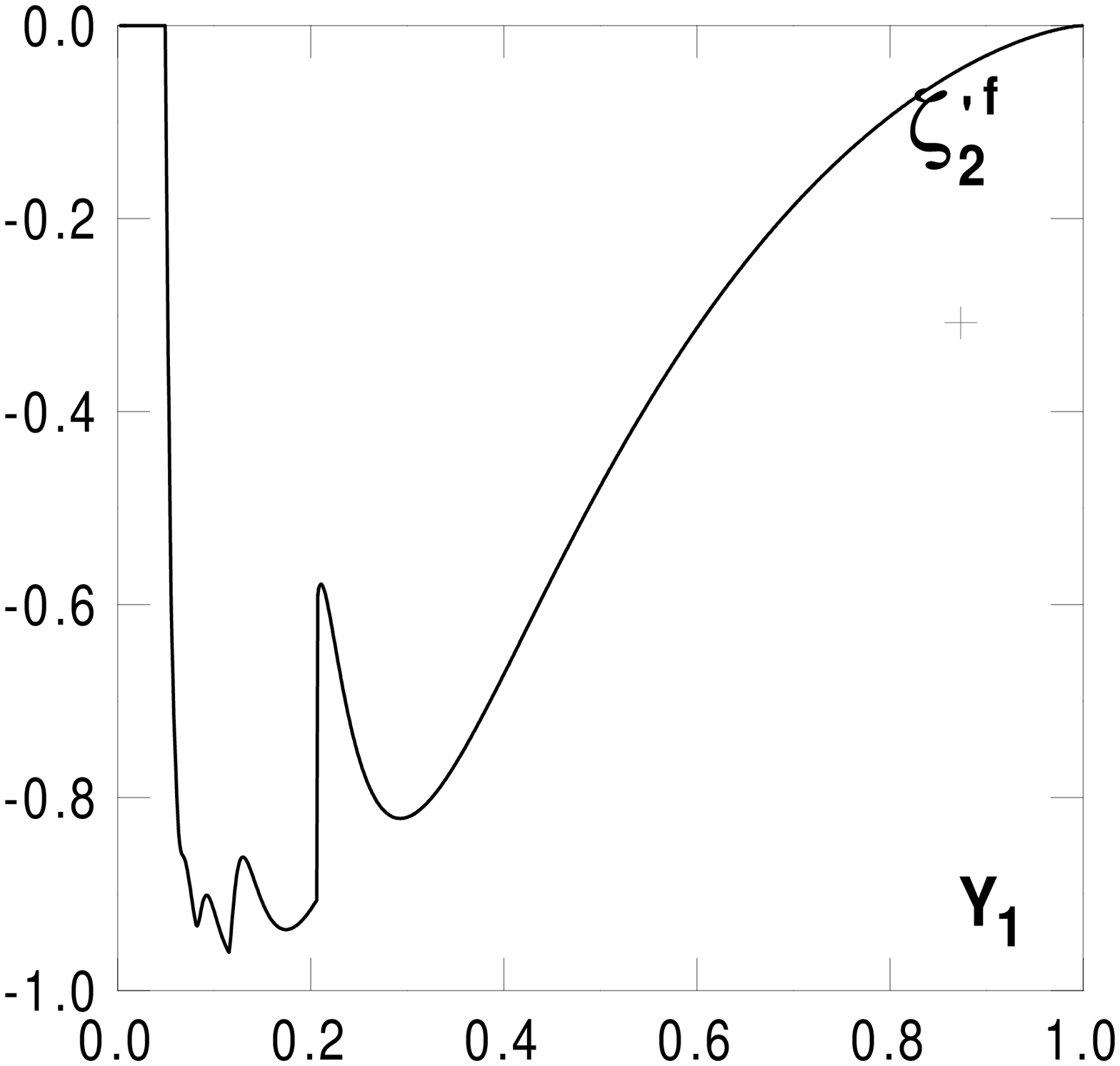}
\vspace{5cm}
\caption{
Components of the polarization vector of the scattered electron
of its energy
$\zeta^{\prime (f)}_2$, $\zeta^{\prime (f)}_3$ at $2\lambda_e=1,\; P_c=+1$;
}
\end{figure}

\begin{figure}
\hspace{2cm}
\includegraphics[bb=110 630 240 770,scale=0.3]{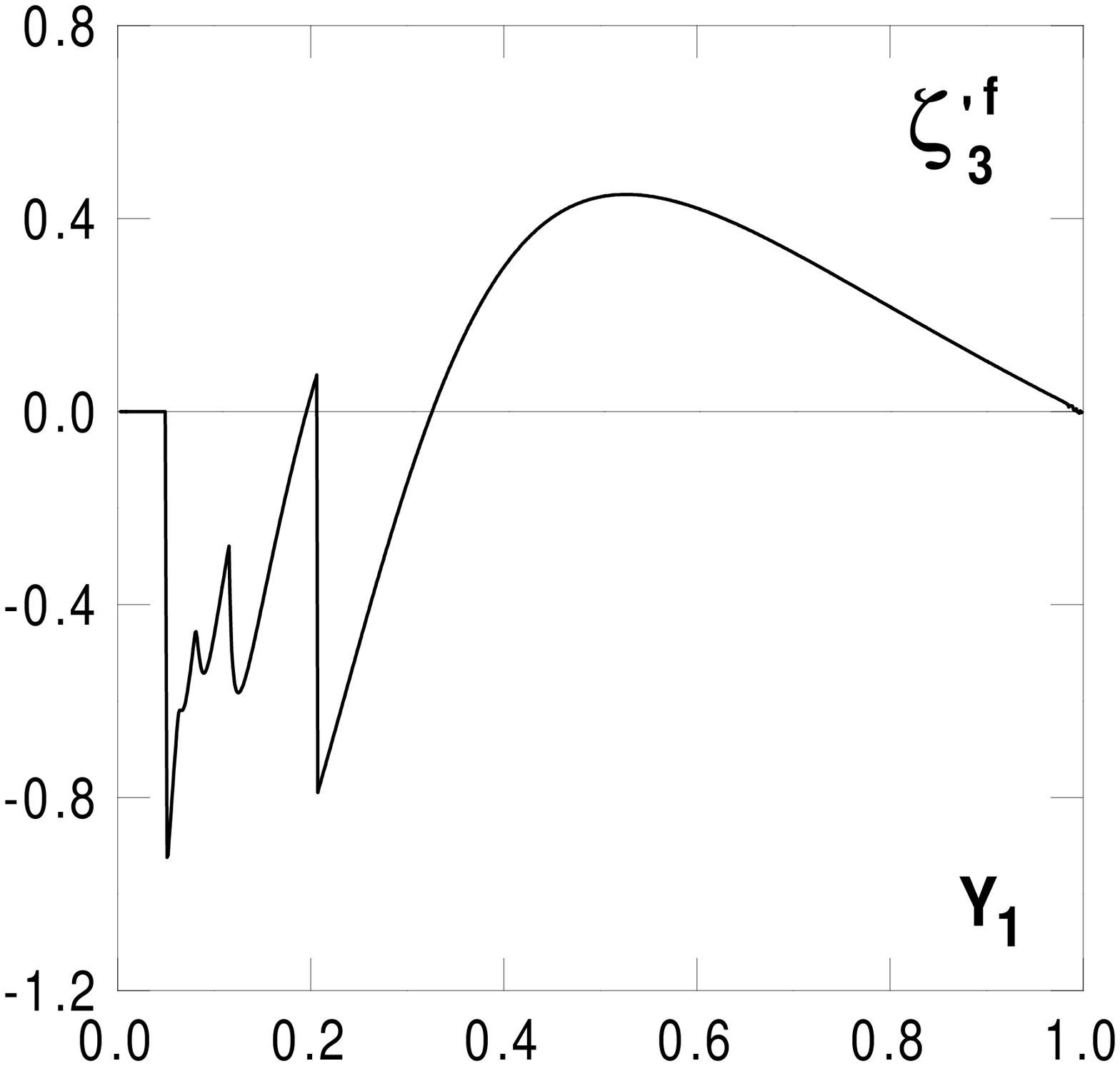}
\hspace{7.5cm}
\includegraphics[bb=340 630 480 770,scale=0.3]{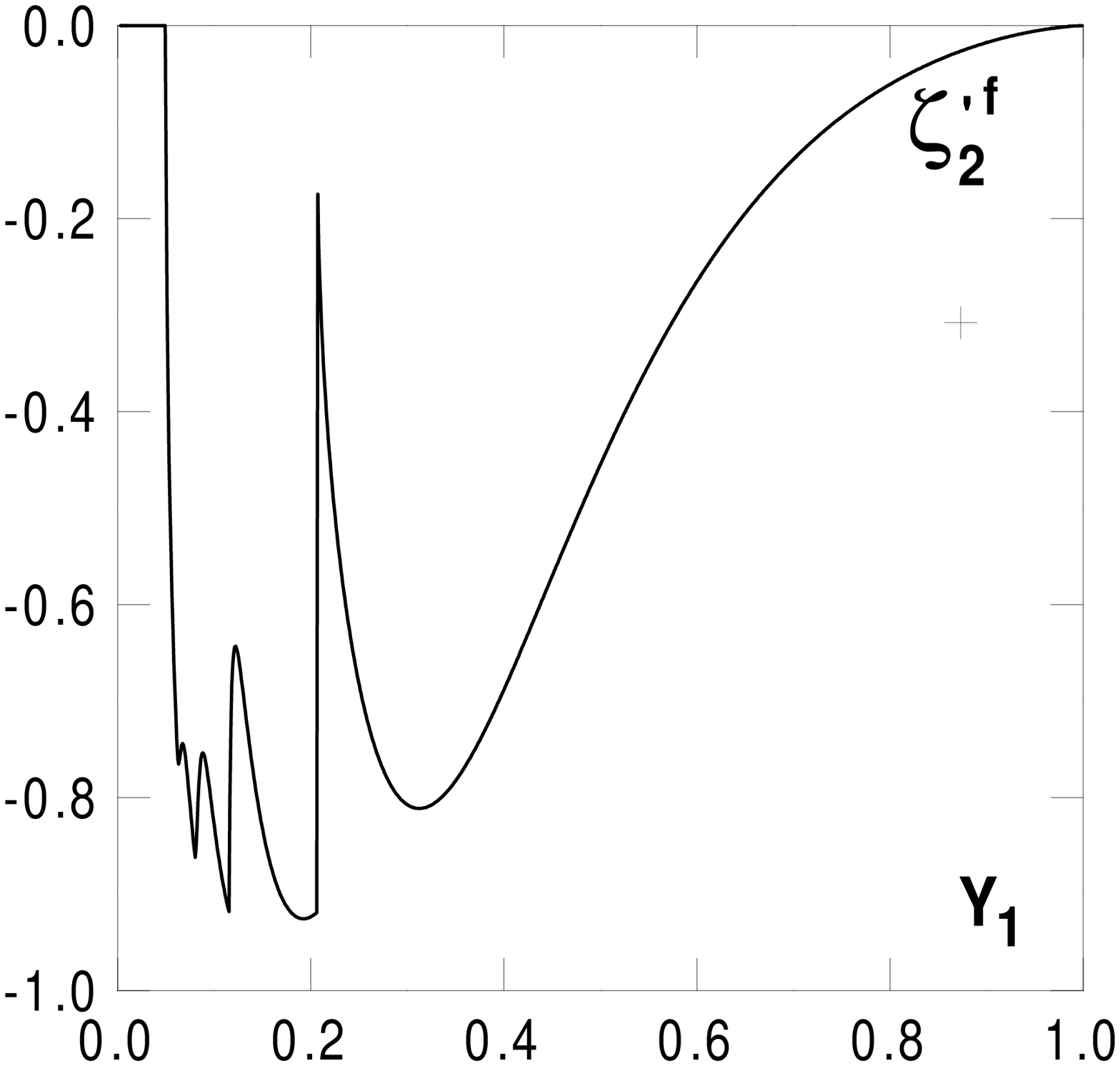}
\vspace{5cm}
\caption{
 Components of the polarization vector of the scattered electron
of its energy
$\zeta^{\prime (f)}_2$, $\zeta^{\prime (f)}_3$ at $2\lambda_e=0,\; P_c=+1$;
}
\end{figure}

\newpage
\vspace{1cm}
\begin{figure}
\hspace{2cm}
\includegraphics[bb=110 630 240 770,scale=0.3]{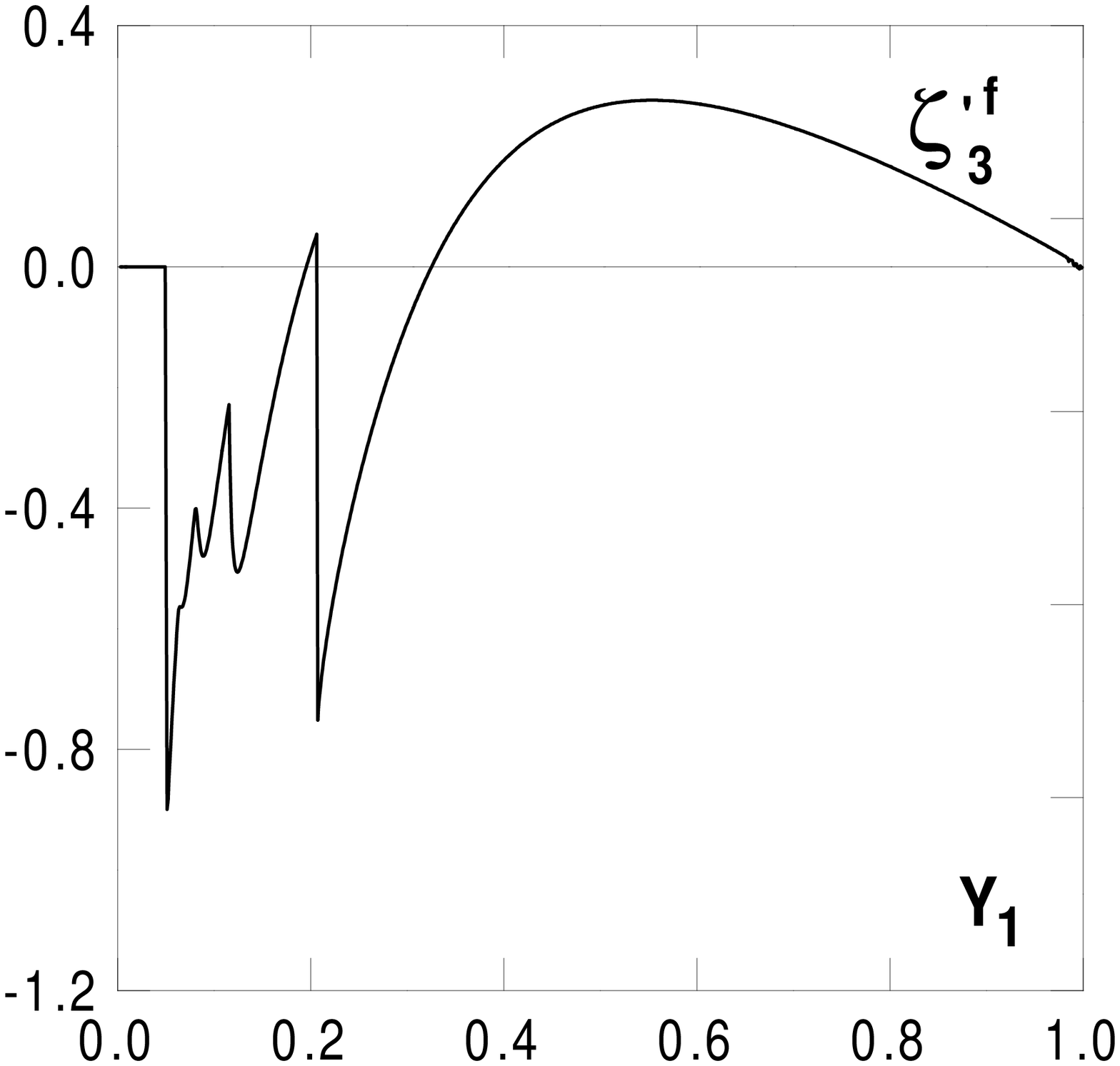}
\hspace{7.5cm}
\includegraphics[bb=340 630 480 770,scale=0.3]{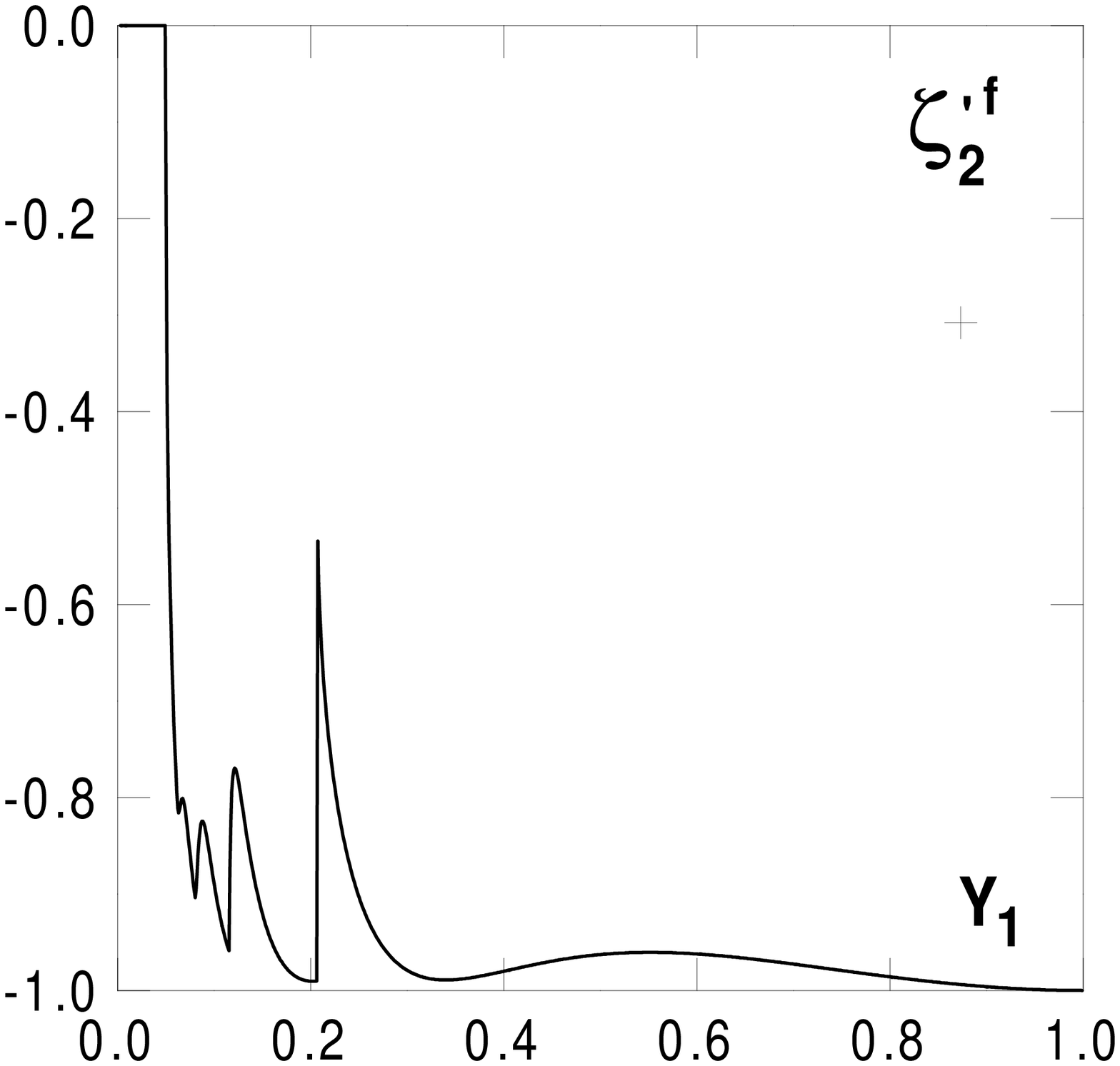}
\vspace{5cm}
\caption{
Components of the polarization vector of the scattered electron
of its energy
$\zeta^{\prime (f)}_2$, $\zeta^{\prime (f)}_3$ at $\zeta_2=-1,\;P_c=1$;
}
\end{figure}

\begin{figure}
\hspace{2cm}
\includegraphics[bb=110 630 240 770,scale=0.3]{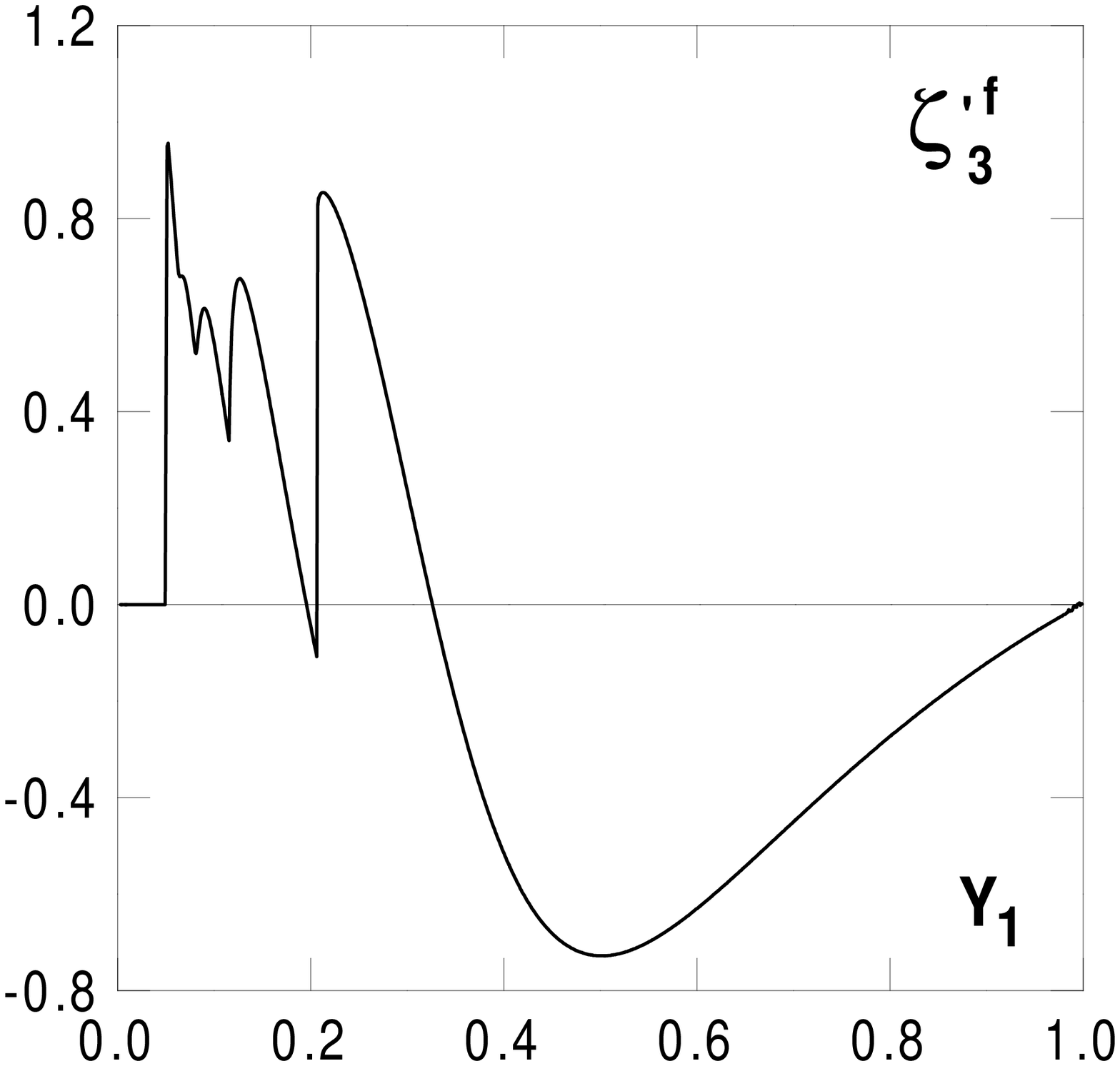}
\hspace{7.5cm}
\includegraphics[bb=340 630 480 770,scale=0.3]{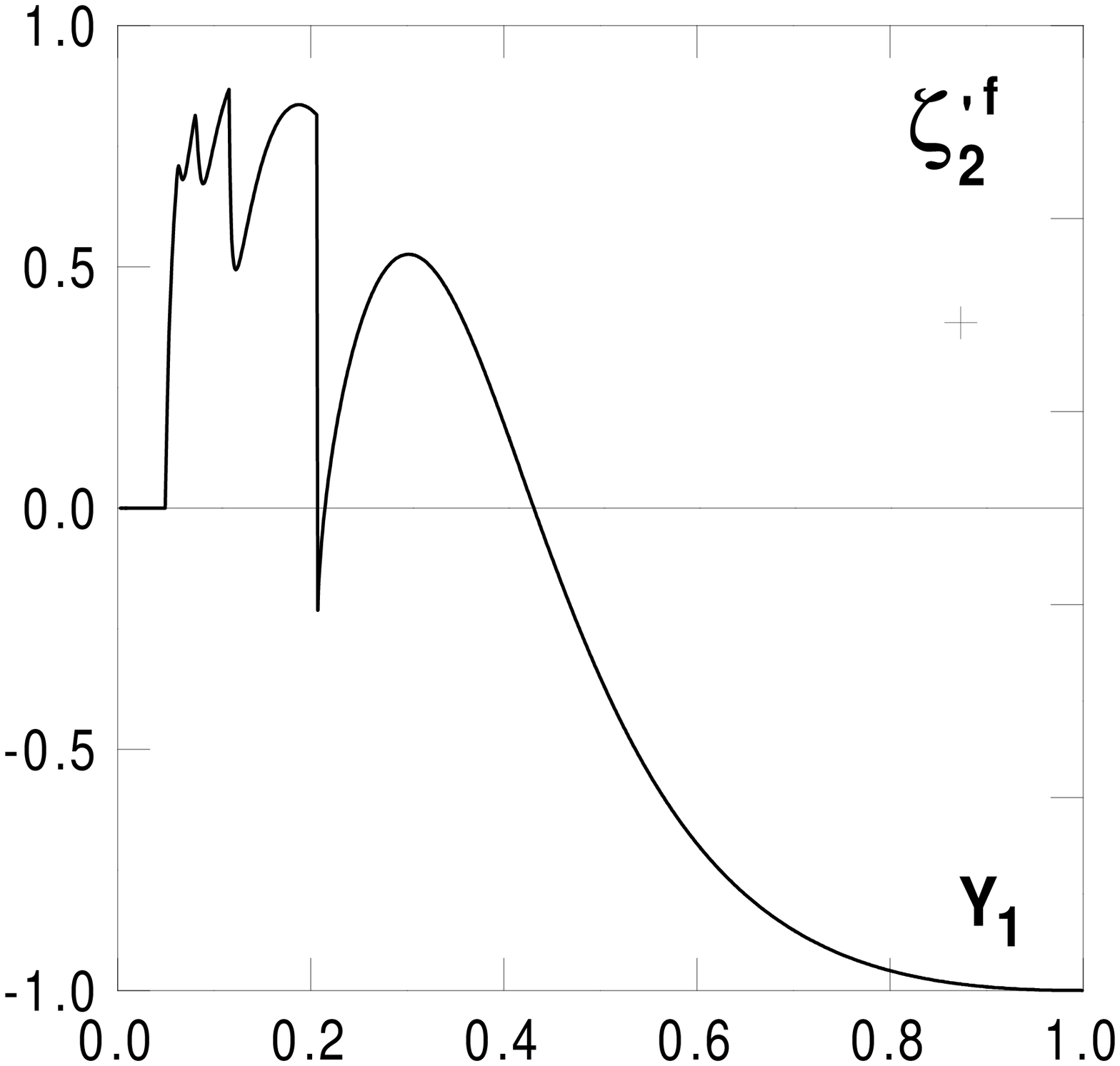}
\vspace{5cm}
\caption{
Components of the polarization vector of the scattered electron
of its energy
$\zeta^{\prime (f)}_2$, $\zeta^{\prime (f)}_3$ at $\zeta_2=-1,\; P_c=-1$;
}
\end{figure}

\begin{figure}
\hspace{2cm}
\includegraphics[bb=110 630 240 770,scale=0.3]{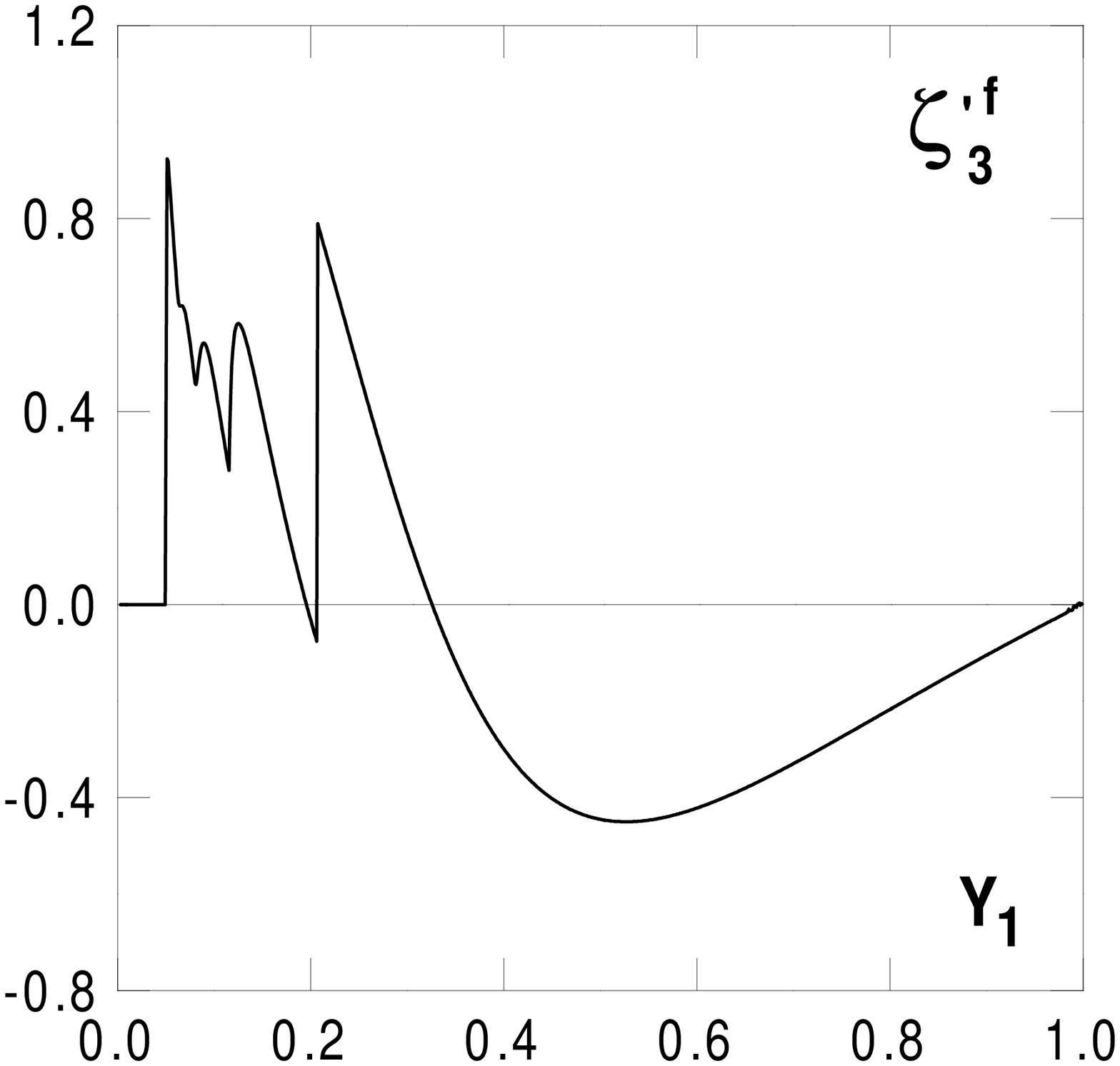}
\hspace{7.5cm}
\includegraphics[bb=340 630 480 770,scale=0.3]{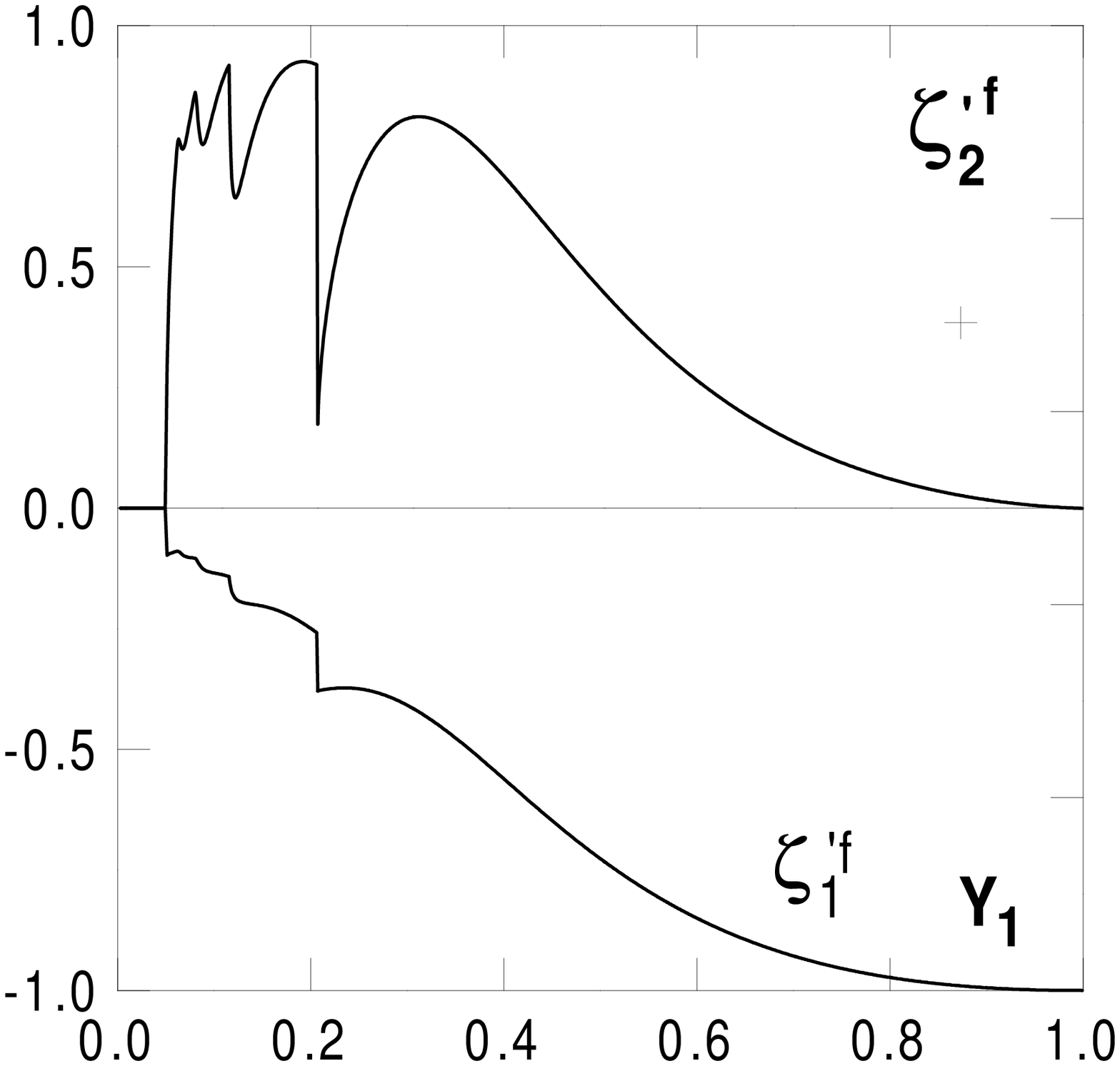}
\vspace{5cm}
\caption{
Components of the polarization vector of the scattered electron
of its energy
$\zeta^{\prime (f)}_2$, $\zeta^{\prime (f)}_3$, $\zeta^{\prime (f)}_3$
at $\zeta_1=-1,\; P_c=-1$;
}
\end{figure}

\end{document}